\documentclass[aps,prl,reprint,showpacs,amsmath,groupedaddress]{revtex4-1}

\usepackage{graphicx}
\usepackage{tabularx}
\usepackage{amsmath}
\usepackage{array}
\usepackage[colorlinks,linkcolor = blue,citecolor = blue,urlcolor = blue]{hyperref}

\begin{document}

\title{Production of Rubidium Bose-Einstein Condensates at a 1 Hz Rate}

\author{Daniel M. Farkas}
\email{daniel.farkas@coldquanta.com}
\author{Evan A. Salim}
\author{Jaime Ramirez-Serrano}
\affiliation{ColdQuanta, Inc., 3030 Sterling Circle, Boulder, CO 80301}

\date{\today}

\begin{abstract}
We present an experimental apparatus that produces Bose-Einstein condensates (BECs) of $^{87}$Rb atoms at a rate of 1\,Hz. As a demonstration of the system's ability to operate continuously, 30 BECs were produced and imaged in 32.1\,s. Without imaging, a single BEC could be produced in 953\,ms. The system uses an atom chip to confine atoms in a dimple trap with frequencies exceeding 1\,kHz. With this tight trap, the duration of evaporative cooling can be reduced to less than 0.5\,s. Using principal component analysis, insight into the largest sources of noise and drift was obtained by extracting the dominant contributions to the variance. The system utilizes a compact physics package that can be integrated with lasers and electronics to create a transportable ultracold-atom device for applications outside of a laboratory environment.
\end{abstract}

\maketitle

While devices based on cold and ultracold neutral atoms offer unprecedented accuracy and precision for a wide variety of applications, the long times needed to cool atoms into the degenerate regime limit the bandwidths that can be achieved from sensors and devices that utilize ultracold matter. A typical laboratory-based apparatus can produce a Bose-Einstein condensate (BEC) in tens of seconds, corresponding to a sensor with a measurement bandwidth less than 0.1\,Hz~\cite{Lewandowski2003,Streed2006}. Compare this, for example, to MEMS-based or laser-based inertial sensors, which typically operate at bandwidths of at least 1\,kHz, high enough to accurately measure rapidly varying signals that are of interest for navigation. For ultracold atoms to compete with existing technologies, the production rates of BECs, quantum gases, and coherent atom sources must improve. Specific applications that would benefit from faster ultracold-atom production include timekeeping~\cite{Hinkley2013,Bloom2013,Maineult2012}, inertial sensing~\cite{Gustavson1997,Gustavson2000,Wang2005,Durfee2006,Burke2009,Berrada2013,McDonald2013}, gravimetry~\cite{Chu1999,McGuirk2002}, electromagnetic field sensing~\cite{Wildermuth2006,Boohi2010}, atomtronics~\cite{Stickney2007,Pepino2009,Ramanathan2011} and space-based experiments and devices~\cite{Ramirez-Martinez2011,Muntinga2013}.

\begin{figure*}[t]
\includegraphics[width=6.5in]{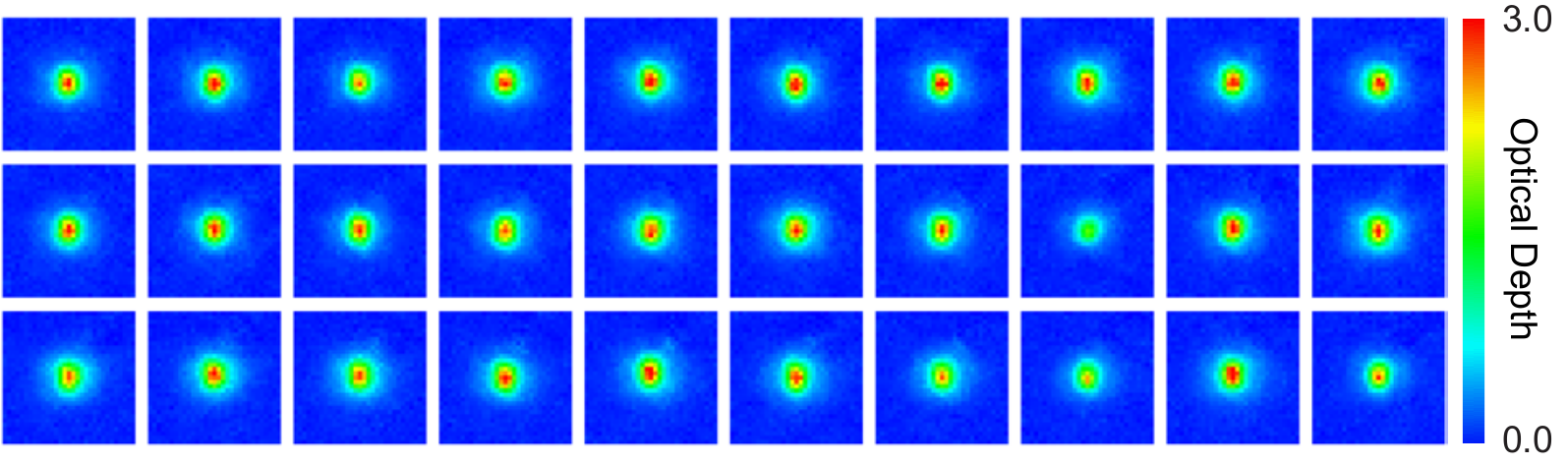}
\caption{Absorption images of 30 $^{87}$Rb BECs produced and imaged in 32.1\,s. Without imaging, a single BEC could be produced in 953\,ms, corresponding to a production rate that exceeds 1\,Hz.}
\label{fig:BECmatrix}
\end{figure*}

As a step towards rapid generation of ultracold matter, we present an apparatus that produces $^{87}$Rb BECs at a rate of 1\,Hz. The system's capabilities are demonstrated in Fig.~\ref{fig:BECmatrix}, which shows 30 BECs produced and imaged in 32.1\,s. Without imaging, a single BEC could be produced in 953\,ms, corresponding to a production rate that exceeds 1\,Hz. The apparatus and techniques described in this work show the extent to which speed improvements can be attained when ultracold matter is generated in a pulsed fashion. Given the engineering challenges associated with developing continuous coherent atomic sources (e.g. an ``atom laser"), pulsed-mode operation remains a viable path toward improving the bandwidths of ultracold-atom sensors~\cite{Mewes1997,Bloch1999,Chikkatur2002,Lahaye2005,Robins2008,Gattobigio2012,Power2012}.

To increase the speed of forced evaporative cooling, which is typically the slowest step in BEC production, atoms are confined in a tight dimple trap created with an atom chip~\cite{Ott2001,Hansel2001,Reichel2002}. The atom chip forms the top wall of a two-chamber ultrahigh vacuum system whose compact size allows for quick loading of a 3D MOT and fast transport of atoms to the atom chip~\cite{Farkas2010,Salim2011}.

\begin{figure}[b]
\includegraphics[width=3in]{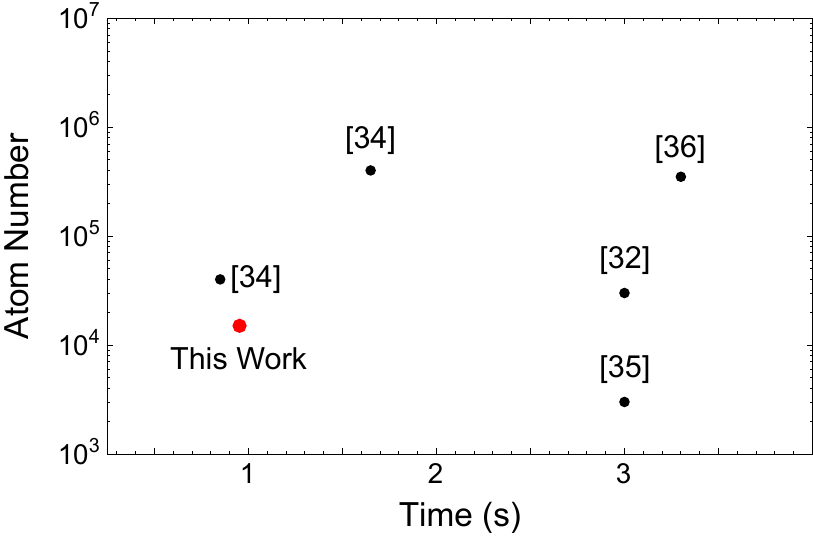}
\caption{Production times and atom numbers for some of the fastest BEC experiments~\cite{Herr,Horikoshi2006,Farkas2010,Kinoshita2005}. The experiment described in this work is one of two to have achieved a BEC production time of less than 1\,s. For Ref.~\cite{Herr}, two data points are shown, one optimized for speed and one optimized for atom number.}
\label{fig:BECmachines}
\end{figure}

As shown in Fig.~\ref{fig:BECmachines}, the apparatus described here is one of the first to have demonstrated BEC production from a hot vapor in less than 1\,s~\cite{Herr,Horikoshi2006,Farkas2010,Kinoshita2005}. All the  experiments represented in this plot used $^{87}$Rb. With the exception of the data point at 3.3\,s, which corresponds to an optical trap~\cite{Kinoshita2005}, all of the experiments used atoms that were magnetically confined with an atom chip.

In addition to speed, ultracold-atom devices must produce samples of ultracold atoms whose proprieties (e.g. number, condensate fraction, position) are stable over time. To that end, principal component analysis was applied to the BEC images in Fig.~\ref{fig:BECmatrix}~\cite{Segal2010}. The characteristic spectrum of eigenvalues obtained from this analysis serves as a baseline for stability and repeatability to which future experiments can be compared. Here, it helped identity sources of noise and drift in the system. For the data set shown in Fig.~\ref{fig:BECmatrix}, most of the variance was attributed to the first three principal components. From plots of these dominant eigenvectors, fluctuations in atom number and the vertical position of the cloud were identified.

The first part of this paper presents an overview of the BEC apparatus, including the vacuum system, atom chip, coils and magnets, physics package, and laser system. Following this is a description of the experimental procedure, including the operating values used to create a BEC in less than 1\,s and the implementation of absorption imaging. To characterize the performance of our vacuum system, data is presented from diagnostic measurements of the 3D MOT lifetime and the magnetic chip-trap lifetime. Finally, the principal component analysis is discussed.

\section{Apparatus}
\subsection{Vacuum Chamber and Atom Chip}

\begin{figure}[b!]
\includegraphics[width=2.95in]{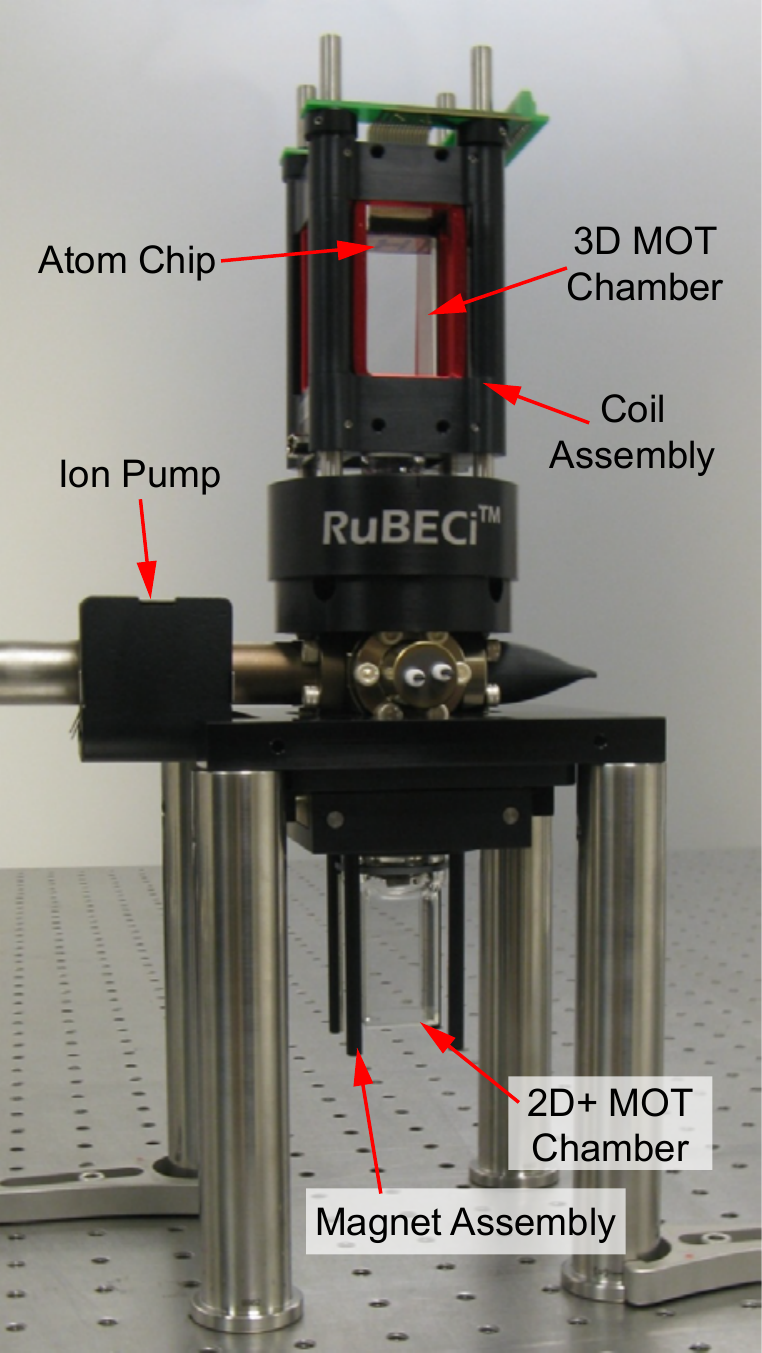}
\caption{The vacuum system consists of two chambers separated by a pinhole. In the lower chamber, a 2D+ MOT creates a cold-atom beam that feeds a 3D MOT in the upper chamber. An atom chip was bonded to the top of the upper chamber, where it forms a wall of the vacuum system. Vacuum is maintained with a 2\,l/s ion pump.}
\label{fig:RuBECi}
\end{figure}

The ultrahigh vacuum system (ColdQuanta RuBECi) utilizes a two-chamber configuration in which a 2D+ magneto-optical trap (MOT) in one chamber feeds a 3D MOT in a second chamber (see Fig.~\ref{fig:RuBECi})~\cite{Farkas2010,Salim2011}. Differential pumping between the chambers is introduced by a 0.75\,mm diameter pinhole in a silicon disc. With this differential pumping, a rubidium pressure of 10$^{-7}$\,Torr can be created in the lower chamber while a 2\,l/s ion pump maintains a pressure below 10$^{-9}$\,Torr in the upper chamber.

A rubidium vapor in the lower chamber is produced by running a current between 3.5 and 3.8\,A through an alkali metal dispenser. A few milliwatts of laser light produces a vertically-oriented ``push" beam that forces atoms through the pinhole and into the upper chamber, where they are collected in a 3D MOT. Reflection of the push beam off the silicon disc in the region around the pinhole produces additional cooling in the third dimension, thereby forming a 2D+ MOT~\cite{Dieckmann1998}.

\begin{figure}[t]
\includegraphics[width=2.6in]{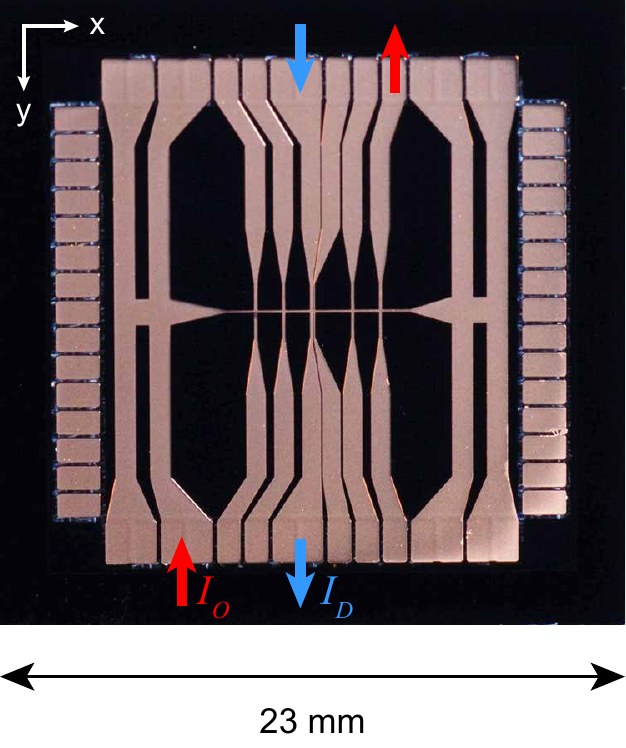}
\caption{The vacuum side of the atom chip consists of copper traces that can produce a variety of Z-traps and dimple traps. The red and blue arrows indicate the traces and currents used for this work. In this photograph, the left side of the atom chip is closest to the ion pump (see Fig.~\ref{fig:RuBECi}).}
\label{fig:UtilityChip}
\end{figure}

Bonded to the top of the upper chamber of the vacuum system is the atom chip shown in Fig.~\ref{fig:UtilityChip}. The atom chip has dimensions of 23 $\times$ 23\,mm and a thickness of 420\,$\mu$m. The substrate contains regions of both glass and silicon, and through-chip vacuum-compatible electrical vias. The vacuum side of the chip contains copper traces that can be driven in different configurations to produce a variety of Z-traps and dimple traps. The traces used in this work have a width of 100\,$\mu$m and a thickness of 10\,$\mu$m.

\begin{figure}[t]
\includegraphics[width=3in]{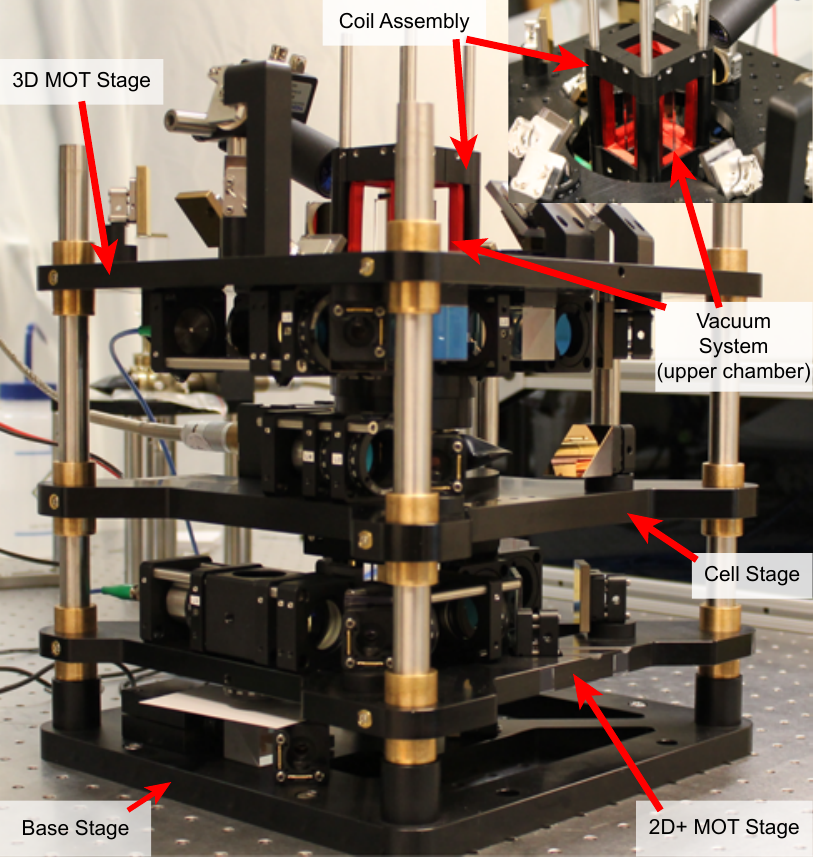}
\caption{Consisting of four stages, the physics package serves as a mounting structure for the vacuum system, free-space optics, magnets, coils, and cameras. The transfer Z-coil and absorption imaging CCD camera are not shown.}
\label{fig:PhysicsPackage}
\end{figure}

\begin{figure}[t]
\includegraphics[width=2.9in]{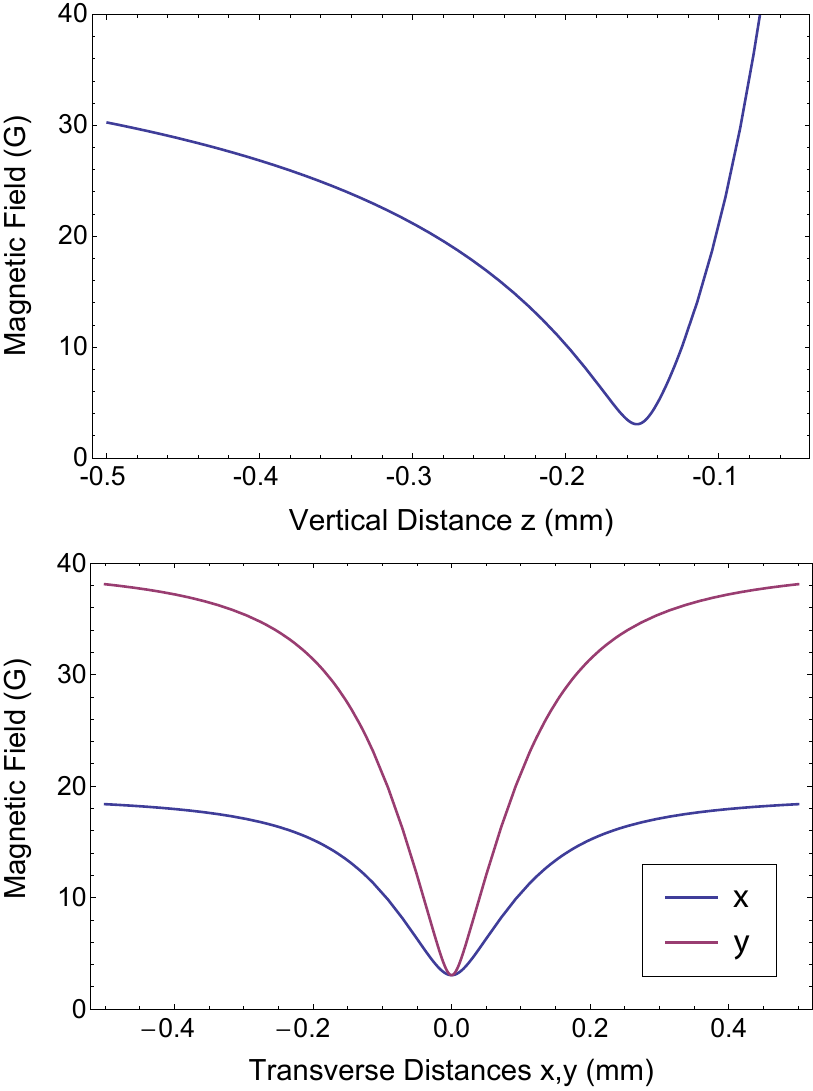}
\caption{To increase the speed of evaporative cooling, a dimple trap with high trap frequencies is used. These plots show the calculated magnitude of the trap's magnetic field along the vertical direction (top) and transverse directions (bottom). The calculated trap frequencies are 2$\pi \times$ (1.96, 1.93, 0.35)\,kHz. The trap minimum of 3.1\,G is located 154\,$\mu$m beneath the chip surface.}
\label{fig:DimpleTrap}
\end{figure}

The duration of evaporative cooling can be reduced to less than 0.5\,s by utilizing a dimple trap with high trap frequencies~\cite{Horikoshi2006,Reichel2001}. This trap is formed by running currents $I_0 = +3.25$\,A and $I_D = +1.3$\,A through two traces that cross each other perpendicularly at the center of the chip. The directions of these currents are represented by the red and blue arrows, respectively, in Fig.~\ref{fig:UtilityChip}. To complete the trap, external bias fields of +19.2\,G and -39.8\,G are applied along the x and y directions, respectively.

The trap parameters were calculated with the Biot-Savart law under the assumption that the chip traces consist of infinitely-thin current sheets having a width of 100\,$\mu$m and various lengths. Figure~\ref{fig:DimpleTrap} shows the magnitude of the magnetic field along the three cardinal directions. The trap minimum of 3.1\,G is located 154\,$\mu$m beneath the chip surface. The calculated trap frequencies are 2$\pi\times$(1.96, 1.93, 0.35)\,kHz.

\subsection{Physics Package}

Shown in Fig.~\ref{fig:PhysicsPackage}, the physics package consists of four stages that can be independently positioned along four stainless steel support rods. These stages serve as mounting baseplates for the vacuum system, free-space optics, magnets, coils, and cameras. At the bottom, the base stage serves as a foundation for the entire package; it contains optics for the 2D+ MOT push beam and a CCD camera for imaging the 2D MOT along the vertical axis.

Above the base stage is the 2D MOT stage. A fiber-optic patchcord delivers MOT laser light (i.e. cooling and repumping light) to this stage, after which free-space optics provide beam shaping, polarization control, and beam splitting. These optics provide the correct beam configuration to produce a 2D+ MOT in the lower chamber of the vacuum system.

The second stage from the top is the cell stage. The center of the vacuum system is bolted to this stage, as well as the permanent magnet assembly used to create the 2D+ MOT. This stage also contains a fiber-optic port, turning mirrors, and waveplates for directing an optical pumping laser beam through the upper chamber of the vacuum system.

On the top level, or 3D MOT stage, laser light delivered by a fiber-optic patchcord is directed with free-space optics to provide the correct beam geometry for producing a 3D MOT in the upper chamber of the vacuum system. This stage also contains a fiber-optic port and turning mirrors that direct an absorption laser beam through the cell and into a CCD camera (Basler Scout 1390fm with Infinity Photo-Optical InfiniMini lens system). The absorption laser beam propagates along the +y direction (using the coordinate system defined in Fig.~\ref{fig:UtilityChip}). To quantify the size and loading rate of the 3D MOT, a biconvex lens images fluorescence photons onto a silicon photodetector (Thorlabs PDA100A)~\cite{Lewandowski2003}.

\subsection{Magnets and Coils}

\begin{figure}[t]
\includegraphics[width=3.2in]{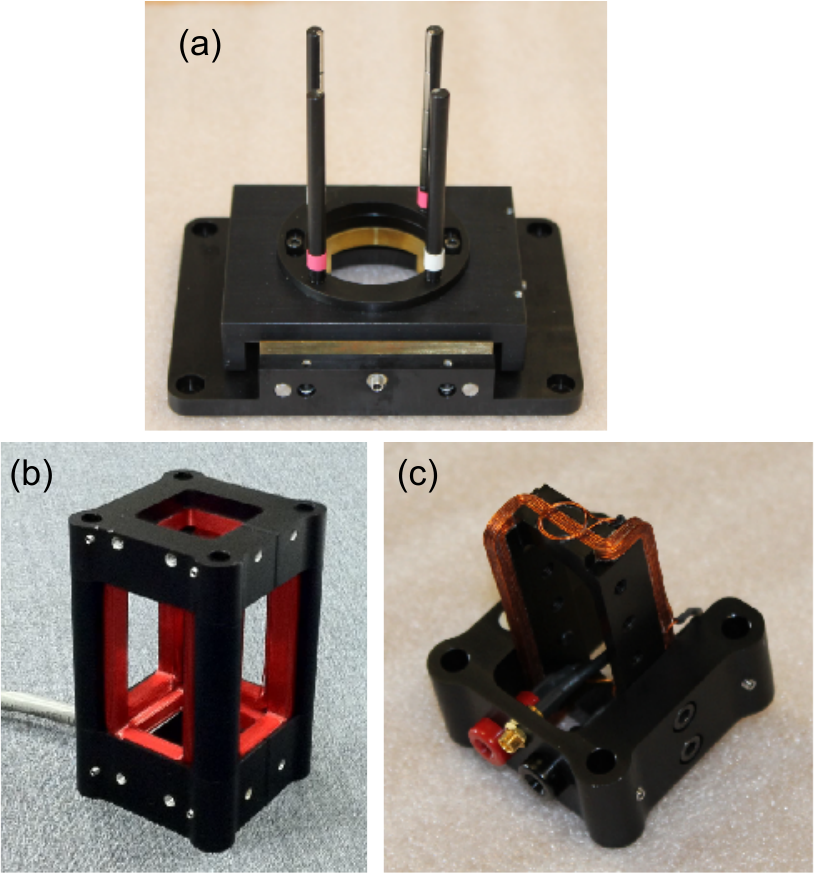}
\caption{(a) The quadrupole field for the 2D+ MOT is generated by a magnet assembly that surrounds the lower chamber of the vacuum system. The magnets are mounted on a stage for translating the position of the MOT axis. (b) Consisting of six coils arranged in a square cuboid, the coil assembly produces magnetic bias fields and gradients in the upper chamber of the vacuum system. (c) Atoms are transported vertically to the atom chip using a transfer ``Z" coil whose windings have the same ``Z" orientation as the chip trap. The circular windings in the center of the coil is an rf loop used for forced evaporative cooling.}
\label{fig:Magnetics}
\end{figure}

The magnetic quadrupole field for the 2D+ MOT is generated by a magnet assembly that surrounds the lower chamber of the vacuum system (see Fig.~\ref{fig:Magnetics}a). To maximize the cold-atom flux driven out of the 2D+ MOT with the vertical push laser beam, the magnets are mounted on a translation stage that is used to align the MOT axis with the pinhole. The magnetic field gradient is approximately 37\,G/cm.

All of the magnetic bias and quadrupole fields needed for the upper chamber are produced by the coil assembly shown in Fig.~\ref{fig:Magnetics}b. This assembly consists of six rectangular coils arranged as a square cuboid. Each of the four rectangular coils on the sides has a nominal DC resistance of 2.4\,$\Omega$ and an inductance of 0.92\,mH (at 1\,kHz). When the two coils aligned along the x-axis are driven with currents of the same magnitude and opposite sign, they produce at the center of the assembly a gradient of (15.1, 11.0, 4.0)\,G/cm/A. When these currents have the same sign, the coils produce a bias field of +22.6\,G/A~$\hat{x}$. The y-axis coils are connected in series to produce a bias field of +22.6\,G/A~$\hat{y}$. The square coils at the top and bottom are connected in series to produce a bias field of +15.6\,G/A~$\hat{z}$. Each of these square coils has a nominal DC resistance of 3.7\,$\Omega$ and an inductance of 1.75\,mH (at 1\,kHz). Measurements of these gradients and bias fields agree with the calculated values listed here to within a few percent.

\begin{figure*}[t]
\includegraphics[width=6.5in]{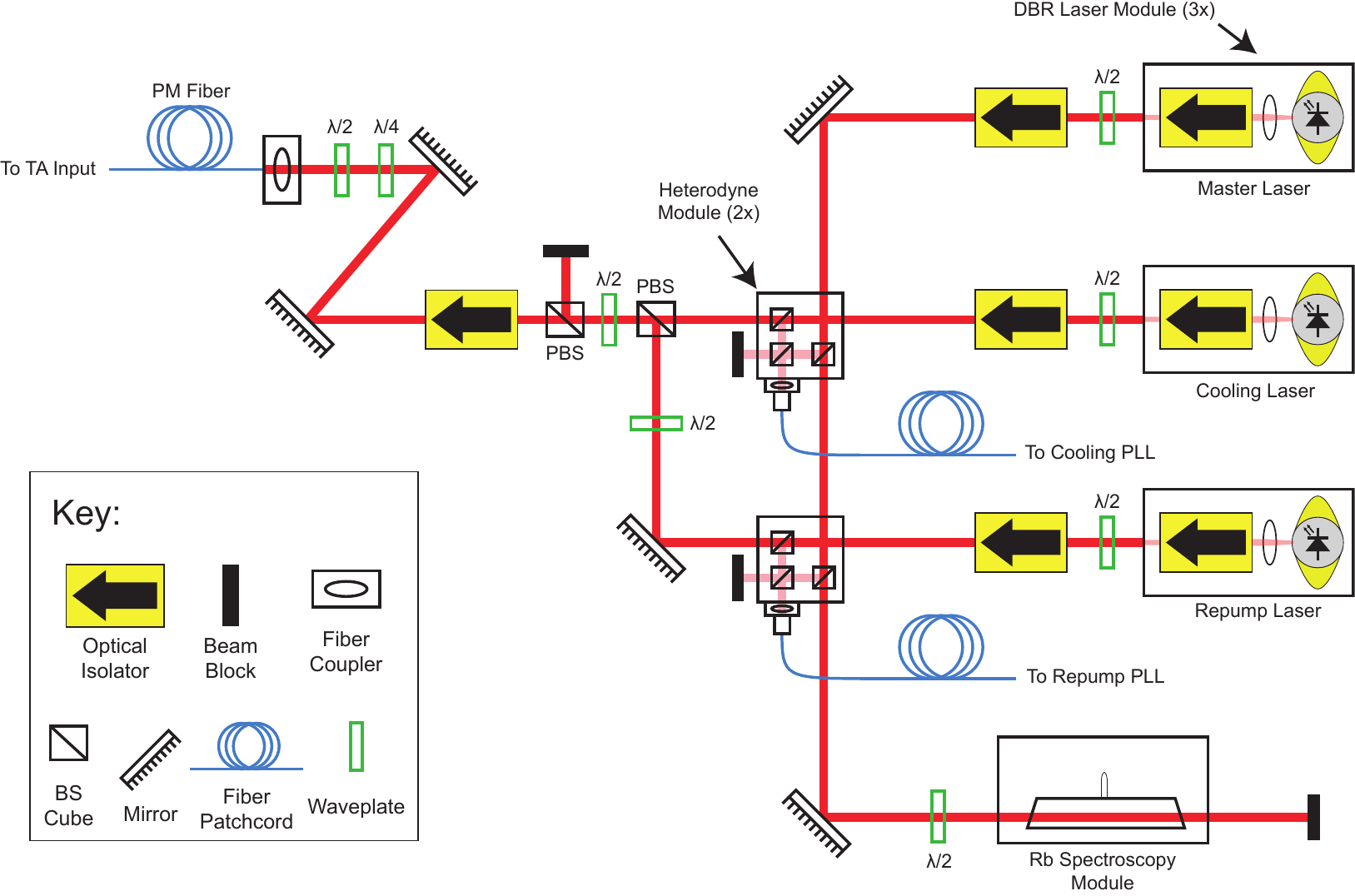}
\caption{The laser system consists of three distributed Bragg reflector (DBR) diode lasers at 780.24\,nm. A master laser locked to a rubidium vapor cell serves as an optical frequency reference whose stability is transferred to the other lasers via optical phase-lock loops. Light at the cooling and repumping transitions is combined into a single beam and amplified up to 200\,mW with a tapered amplifier (TA).}
\label{fig:LaserSystemBeforeTA}
\end{figure*}

Cold atoms are transported vertically from the center of the 3D MOT cell to the atom chip using a transfer ``Z" coil (see Fig.~\ref{fig:Magnetics}c). In conjunction with external bias fields, this coil creates a Ioffe-Pritchard trap whose height varies with current. This coil has a nominal DC resistance of 0.7\,$\Omega$ and an inductance of 110\,$\mu$H (at 1\,kHz). Affixed to the bottom of the coil is an rf loop for evaporative cooling. The coil is mounted above the vacuum system such that the rf loop is situated less than 1\,mm above the ambient side of the atom chip.

\subsection{Laser System}

\begin{figure*}[t]
\includegraphics[width=6.5in]{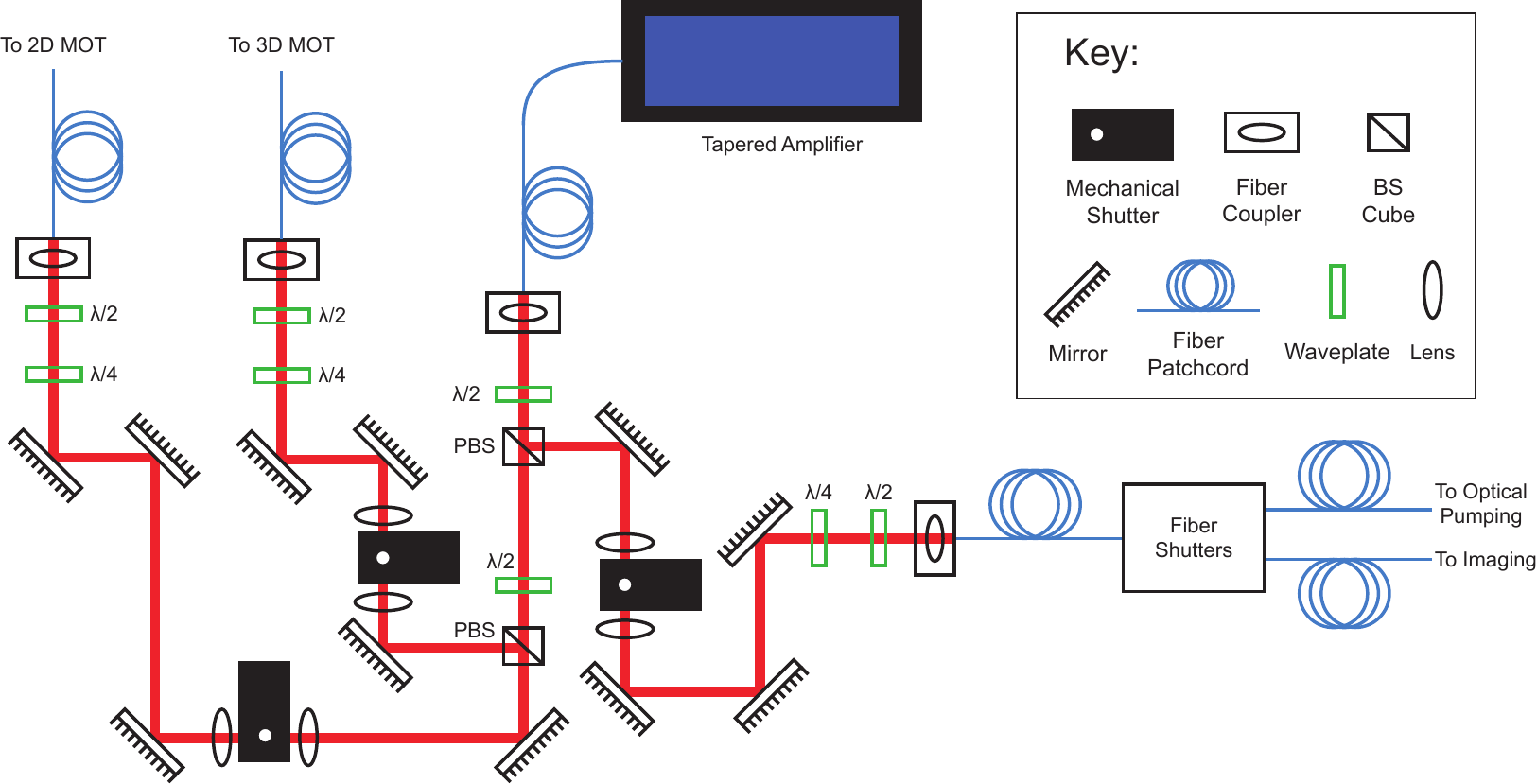}
\caption{The 200\,mW output of the tapered amplifier is split into three beams. All patchcords shown use PM fiber and FC/APC connectors. The 2D MOT and 3D MOT patchcords output 72\,mW and 50\,mW, respectively, of combined cooling and repumping power. For optical pumping and imaging, a pair of fiber-optic shutters outputs 0.5\,mW.}
\label{fig:LaserSystemAfterTA}
\end{figure*}

The laser system uses three distributed Bragg reflector (DBR) diode lasers (Vescent Photonics D2-100-DBR-780) operating near the rubidium D$_2$ line at 780.24\,nm. As shown in Fig.~\ref{fig:LaserSystemBeforeTA}, the stability of a frequency-stabilized master laser is transferred to the other lasers via optical phase-locked loops (PLLs). The frequency of the master laser is locked to the $F=3 \rightarrow F'=3,4$ crossover resonance in $^{85}$Rb using a Doppler-free saturated absorption spectrometer (Vescent Photonics D2-110-Rb) and servo electronics (Vescent Photonics D2-125-PL-T). To generate a zero-crossing error signal suitable for frequency stabilization, 4\,MHz frequency sidebands are imposed on the master laser via modulation of the laser current. The cooling laser is phase-locked approximately 1.5\,GHz to the red of the master laser, placing its frequency near the $^{87}$Rb $F=2 \rightarrow F\text'=3$ cooling transition. Similarly, the repumping laser is phase-locked 4\,GHz to the blue of the master laser, placing its frequency at the $^{87}$Rb $F=1 \rightarrow F'=2$ repumping transition.

Microwave beat notes between the lasers are generated using heterodyne modules (Vescent Photonics D2-150) that couple a few percent of the light from each beam into multimode fiber-optic patchcords. Each fiber delivers light to a high-speed photodetector that outputs a signal proportional to the difference frequency of the lasers. This difference frequency is divided by a prescaler; the phase of this lower frequency signal is then compared to a local oscillator using a digital phase/frequency detector (Vescent Photonics D2-135-FC-800) whose output is fed back to the laser current controller (Vescent Photonics D2-105). The local oscillator is an rf voltage-controlled oscillator tuned with an external analog voltage.

After passing through their respective heterodyne modules, the outputs of the cooling and repumping lasers are overlapped using a half-waveplate and polarized beamsplitter cube (PBS). A second half-waveplate and PBS set the ratio of cooling and repumping powers and project the two components onto the same polarization axis.  This dual-frequency beam is then coupled into a polarization-maintaining (PM) patchcord that delivers this light to the input of a tapered amplifier (TA; New Focus TA-7613-P). Approximately 15\,mW of light seeds the TA. Although the TA can output up to 500\,mW, it is operated such that it outputs only 200\,mW.

In addition to a 40\,dB optical isolator housed inside each laser module, an additional 40\,dB isolator (Optics For Research IO-3D-780-VLP) is placed after each laser to block reflections originating from the output faces of the multimode patchcords. Reflections from the fiber input of the TA are minimized with the use of FC/APC connectors. However, when the TA is turned on, back-reflected light into the input fiber can cause multimode operation. To block this light, an additional 40\,dB optical isolator is placed in front of the fiber coupler.

As shown in Fig.~\ref{fig:LaserSystemAfterTA}, the TA output is split into three beams: one for the 2D MOT, one for the 3D MOT, and one for both optical pumping and imaging. To ensure the high extinction ratios needed for BEC production, each beam path contains a mechanical shutter (Stanford Research Systems SR475 or SR476). The transition time of each shutter is decreased to less than 100\,$\mu$s by placing it at the center of a 1:1 telescope formed from a pair of $f$=30\,cm plano-convex lenses. Each of the three beams is then coupled into a PM fiber-optic patchcord that delivers the light to the physics package. For the pumping/imaging beam, transition times less than 1\,$\mu$s are achieved with a pair of fiber-optic shutters connected in series (Agiltron SWDR-112211112 and NSSW-127115333). The 2D MOT and 3D MOT patchcords deliver 72\,mW and 50\,mW, respectively, of combined cooling and repumping power to the experiment, while the fiber shutters output 0.5\,mW of power for either optical pumping or imaging.

To minimize reflections, all of the PM patchcords use FC/APC connectors. To ensure the long-term stability of the polarization at the output of each fiber, a half-waveplate and quarter-waveplate align the laser polarization with the fiber axis.

\section{BEC Production}

In this section, the experimental procedure for producing BECs is summarized. As shown in Table~\ref{tab:ExpParameters}, the sequence is divided into a series of stages, each with its own values for laser parameters, magnetic fields, chip currents, and RF drives.

\newcolumntype{C}{>{\centering\arraybackslash}X}
\begin{table*}[t]
	\begin{tabularx}{\textwidth}{p{2.8cm}|C|CC|CCCC|C|CC|C}
			\hline \hline
            \noalign{\smallskip}

            \multicolumn{1}{c|}{\raisebox{-2ex}[0pt]{Stage}}
            & \raisebox{-1.5ex}[0pt]{Time} & \raisebox{-1.5ex}[0pt]{$\Delta_c$} & \raisebox{-1.5ex}[0pt]{$\Delta_r$}
            & \raisebox{-1.5ex}[0pt]{$B_{x1}$} & \raisebox{-1.5ex}[0pt]{$B_{x2}$} & \raisebox{-1.5ex}[0pt]{$B_y$} & \raisebox{-1.5ex}[0pt]{$B_z$}
            & \raisebox{-1.5ex}[0pt]{Z-Coil} & \raisebox{-1.5ex}[0pt]{$I_0$} & \raisebox{-1.5ex}[0pt]{$I_D$} &  RF Freq. \\

            & (ms) & \multicolumn{2}{c|}{(MHz)} & \multicolumn{4}{c|}{(A)} & (A) & \multicolumn{2}{c|}{(A)} & (MHz) \\

	        \noalign{\smallskip}
            \hline
            \noalign{\smallskip}

            3D MOT Loading      & 125 & -15 & 0 & 0.58 & -0.5 & -0.085 & 0.36 & & & & \\
            Compression         & 25 & \hspace{-0.12in}$\rightarrow$~-27 & \hspace{-0.17in}$\rightarrow$~140 & \hspace{-0.15in}$\rightarrow$~3 & \hspace{-0.17in}$\rightarrow$~-2.2 & \hspace{-0.15in}$\rightarrow$~-0.17 & \hspace{-0.15in}$\rightarrow$~0.1 & & & & \\
            PGC                 & 3.32 & -110 & 0 & \multicolumn{2}{c}{-0.002} & -0.014 & 0.046 & & & & \\
            Optical Pumping     & 0.85 & -235 & 0 & \multicolumn{2}{c}{-0.1} & 0 & 0 & & & & \\
            Z-Coil Loading      & 3 & & & \multicolumn{2}{c}{-0.3} & -0.5 & 0.5 & 19.5 & & \\
            Z-Coil Transport    & 210 & & & \multicolumn{2}{c}{\hspace{-0.15in}$\rightarrow$~0.25} & \hspace{-0.15in}$\rightarrow$~-1.80 & \hspace{-0.17in}$\rightarrow$~0 & \hspace{-0.15in}$\rightarrow$~7 & & & \\
            Chip Transfer       & 85 & & & \multicolumn{2}{c}{\hspace{-0.15in}$\rightarrow$~0.85} & \hspace{-0.15in}$\rightarrow$~-1.76 & 0 & \hspace{-0.15in}$\rightarrow$~0 & 3.25 & 1.3 & \\
            Evaporation A       & 110 & & & \multicolumn{2}{c}{\hspace{0.01in}0.85} & -1.76 & 0 & & 3.25 & 1.3 & 35~$\rightarrow$~20 \\
            Evaporation B       & 120 & & & \multicolumn{2}{c}{\hspace{0.01in}0.85} & -1.76 & 0 & & 3.25 & 1.3 & $\rightarrow$~14 \\
            Evaporation C       & 130 & & & \multicolumn{2}{c}{\hspace{0.01in}0.85} & -1.76 & 0 & & 3.25 & 1.3 & $\rightarrow$~5 \\
            Evaporation D       & 90 & & & \multicolumn{2}{c}{\hspace{0.01in}0.85} & -1.76 & 0 & & 3.25 & 1.3 & $\rightarrow$~2 \\
            Evaporation E       & 25 & & & \multicolumn{2}{c}{\hspace{0.01in}0.85} & -1.76 & 0 & & 3.25 & 1.3 & $\rightarrow$~1.14 \\
            Decompression       & 20 & & & \multicolumn{2}{c}{\hspace{-0.1in}$\rightarrow$~0.283} & \hspace{-0.1in}$\rightarrow$~-0.587 & 0 & & 3.25 & 1.3 & \\
            Imaging - Atoms     & 5.28 & 0 & 0 & \multicolumn{2}{c}{0} & 0 & 0 & & & \\
            Imaging - Back.     & 60 & -15 & 0 & \multicolumn{2}{c}{0} & 0 & 0 & & & \\

            \noalign{\smallskip}
			\hline \hline

	\end{tabularx}
\caption{Experimental parameters used to produce BECs at a rate of 1\,Hz. An arrow indicates a linear ramp; the arrow points to the ending value while a blank space to the left of an arrow indicates that the initial value of the ramp is equal to the value of that parameter at the end of the previous stage. The detunings of the cooling and repumping lasers are denoted by $\Delta_c$ and $\Delta_r$; blank spaces in these two columns indicate that the laser beams are shuttered. All magnetic fields are expressed in terms of drive currents and can be converted into magnetic field units using the numerical factors listed in the text. The coil labeled ``x1" is located on the side of the vacuum system closest to the ion pump. Currents through the main and dimple traces of the atom chip are denoted by $I_0$ and $I_D$, respectively. Blank spaces in the Z-coil and chip current columns indicate no current, while blank spaces in the RF frequency column indicate that the frequency generator is disabled. The duration of the first imaging stage includes a 5\,ms time-of-flight.}
\label{tab:ExpParameters}
\end{table*}

The BEC production sequence starts by loading approximately $5\times10^8$ $^{87}$Rb atoms into a 3D MOT that is centered in the upper chamber of the vacuum system. These atoms are loaded from the 2D+ MOT produced in the lower vacuum chamber. The 2D+ MOT is generated with 72\,mW of combined cooling and repumping laser power, of which a few milliwatts is split off to form the vertically oriented push beam~\cite{Dieckmann1998}. The 3D MOT is generated with 50\,mW of combined laser power. For both MOTs, the laser beams are retroreflected, effectively doubling the amount of laser power. The two MOTs are created with light from the same lasers, and therefore both have the same cooling laser detuning of -2.5\,$\Gamma$, where $\Gamma$ = 6.067\,MHz is the natural linewidth of the rubidium D$_2$ transition.

To make this stage as short as possible, the cold-atom flux produced by the 2D+ MOT is maximized by running the rubidium dispenser at 3.8\,A. At this current, the 3D MOT loading rate is typically $4\times10^9$ atoms/s. Operating the dispenser at higher currents was not found to increase the 3D MOT loading rate.

After loading, the 3D MOT is spatially compressed by decreasing the cooling laser detuning to -4.5\,$\Gamma$ and increasing the magnetic field gradient by a factor of 4.8~\cite{Petrich1994}. Atoms accumulate in the dark $F=1$ ground hyperfine state by blue-detuning the repumping laser by 140\,MHz. During this stage, the atom cloud is moved upward by 1 to 2\,mm by changing the vertical bias field. This places the atoms closer to the atom chip, reducing the current needed to initially trap the atoms with the Z-coil.

The atoms are cooled below the Doppler limit with a few milliseconds of polarization gradient cooling~\cite{Dalibard1989}. Here, the cooling laser is red-detuned by 110\,MHz while the repumping laser is on-resonance. Small magnetic bias fields cancel stray fields, particularly those from the nearby ion pump magnets. The cloud's temperature is typically lowered to between 25 and 50\,$\mu$K, as determined from resonant absorption imaging of the cloud size for various expansion times.

To maximize the number of atoms that are magnetically trapped and transported to the atom chip, the atoms are optically pumped into the magnetically trappable $|F=2, m_F=+2 \rangle$ state. Optical pumping typically increases the number of trapped atoms by a factor of 2.5. Here, a magnetic bias field is applied along the x direction. Approximately 0.5\,mW of optical pumping light passes through a quarter-waveplate to become circularly polarized, and propagates through the cell along the -x direction. The cooling laser is red-detuned by 235\,MHz, placing its frequency approximately 5.2\,$\Gamma$ to the blue of the $F=2 \rightarrow F'=2$ transition.

The remaining stages of BEC production require no laser light, and all of the laser beams are now mechanically shuttered to prevent stray photons from heating the atoms and inhibiting condensation. The optically pumped atoms are transported vertically to the atom chip using a magnetic trap created by the external Z-coil and bias coils. With almost 20\,A of current flowing through the Z-coil, the position of the initial trap coincides with the cloud. Linearly reducing this current moves the trap position upward. To prevent excess heating of the atoms during transport, the ramp is made adiabatic by extending its duration to more than 200\,ms.

\begin{figure*}[t]
\includegraphics[width=6.4in]{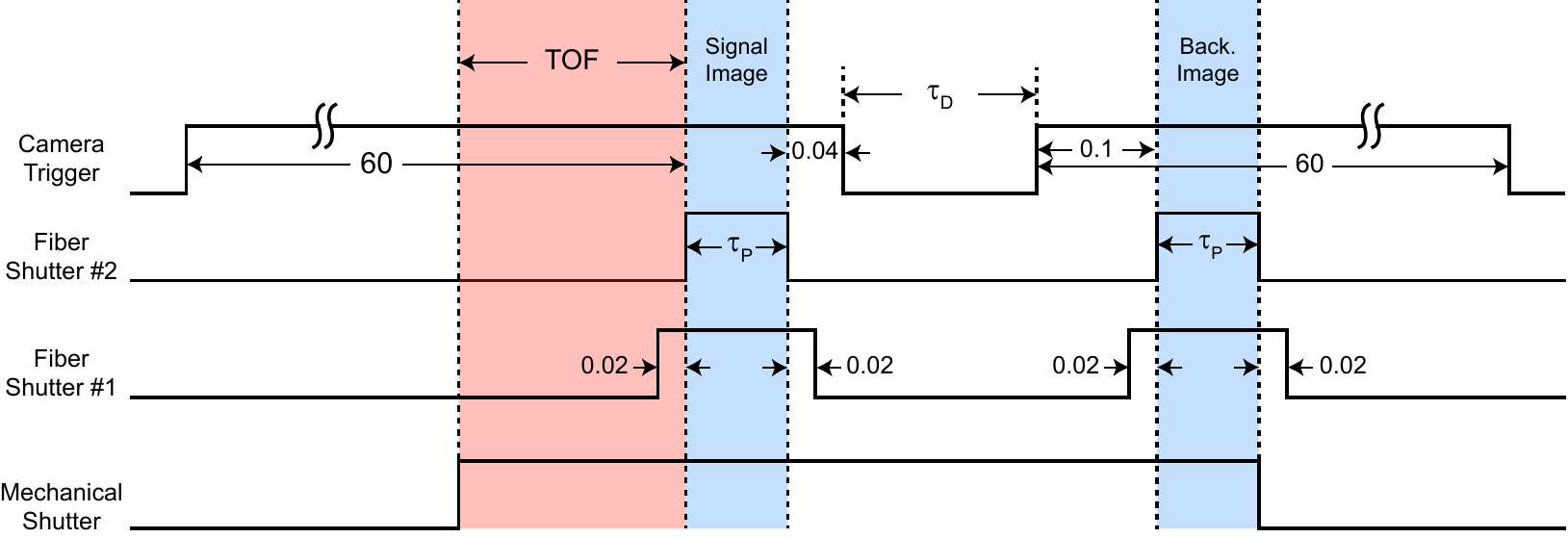}
\caption{Timing diagram for implementing overlapped exposure mode of the CCD camera. By taking the signal image at the end of the first 60\,ms exposure, and the background image at the beginning of the second 60\,ms exposure, the delay time $\tau_D$ between images can be reduced to 0.2\,ms. Here, $\tau_P$ = 0.04\,ms is the duration of the absorption laser pulses. Acquisition of the signal and background images is represented by the blue shading. The time-of-flight (TOF) is represented by the red shading. To account for latency, the background image starts 0.1\,ms after the trigger. All durations listed are in milliseconds. The first imaging stage in Table~\ref{tab:ExpParameters} starts with the time-of-flight.}
\label{fig:ImagingTimingDiagram}
\end{figure*}

Atoms are transferred from the Z-coil trap into the chip trap by ramping the Z-coil current to 0\,A. At the beginning of this transfer stage, the atom chip currents are turned on to their final values of 3.25\,A and 1.3\,A for the main trace and dimple trace, respectively. Evaporative cooling is then implemented with a series of linear frequency ramps from 35 to approximately 1\,MHz~\cite{Ketterle1996}. As shown in Table~\ref{tab:ExpParameters}, the total duration of the five ramps is 475\,ms. The RF drive power is set to 0.07, 0.28, 0.57, 0.57, and 0.11\,W (rms into a 50\,$\Omega$ load) for stages A through E, respectively.

After the cloud condenses, the chip trap is adiabatically decompressed by reducing the x and y bias fields to 1/3 of their nominal values. Decompressing the trap reduces the condensate's chemical potential. The smaller potential energy gradient reduces the rate at which the condensate expands when dropped. Slowing down this expansion makes it easier to visually identify condensed atoms from thermal atoms (which expand at a rate determined solely by their temperature).

\section{Imaging}

Pictures of atoms are obtained using resonant absorption imaging. Two exposures are taken for each cloud. The first, or ``signal", exposure is taken after the cloud has fallen under gravity for a time-of-flight (TOF) between 2 and 12\,ms. The second, or ``background", exposure is taken without the atoms.

Since the absorption laser beam is derived from the TA output, it contains two frequencies. Constituting 90\% of the total power, the dominant component is resonant with the $F=2 \rightarrow F'=3$ transition; scattering of these photons by the atoms is used to measure absorption. The remaining 10\% of the laser beam is repumping light. Although scattering of repumping light is assumed to be negligible, it still contributes to both signal and background exposures. To correct for its presence, the optical depth $\text{OD}_i$ of pixel $i$ is defined as
\begin{eqnarray}
\text{OD}_i=\ln\left(\frac{1-R}{s_i/b_i-R}\right),
\end{eqnarray}
where $s_i$ is the value of the pixel in the signal exposure, $b_i$ is the value of the same pixel in the background exposure, and the repumping component constitutes a fraction $R = 0.1$ of the total laser power. A complete OD image is obtained by repeating this calculation for all 1392 $\times$ 1040 pixels.

The duration of the laser pulses is typically 0.04\,ms, a value that represents a trade-off between competing necessities: on the one hand, the image duration should be as short as possible in order to capture the atoms' instantaneous positions and to prevent smearing due to their motion. At the same time, the duration should be long enough to ensure that the brightest pixels register values close to, but not exceeding, their maximum value of 2$^{12}$ = 4096; this ensures the camera's full dynamic range is utilized. Taken together, these two constraints would lead one to conclude that short, intense laser pulses are ideal. However, the laser intensity should not saturate the atomic transition, as the above formula for OD does not account for this nonlinear effect.

The atoms are imaged without an external magnetic field, and therefore the Zeeman sublevels of each hyperfine level are degenerate. In addition, the absorption laser beam is $\pi$-polarized. For this scenario, the saturation intensity for the $F=2 \rightarrow F'=3$ transition is 3.0538\,mW/cm$^2$~\cite{Steck,Gao1993}. Approximately 0.4\,mW of laser power at the absorption frequency reaches the atoms, and the laser beam has a 1/$e^2$ diameter of 1.5\,cm. The resulting peak intensity is 0.45\,mW/cm$^2$, corresponding to a saturation parameter of $S = 0.15$. For these values, the camera pixels register values that approach, but do not exceed, their maximum value for a pulse duration of 0.04\,ms. The camera is programmed to provide no additional internal gain.

Ideally, optical artifacts will appear identically in both signal and background exposures, and therefore will cancel in the OD image. However, mechanical vibrations can cause these artifacts to shift by several pixels in the interim between the two exposures. This results in imperfect cancellation that usually manifests itself as visible striations (i.e ridges and valleys) in the OD images. To minimize this effect, the delay between the exposures should be made as short as possible, and preferably less than the period of a typical acoustic vibration (e.g. $\ll$\,1\,ms). For commercial CCD cameras, the minimum time delay between back-to-back exposures is usually limited by the speed at which data is transferred from the camera to a computer. The CCD camera used here utilizes a FireWire 800 link, which can upload a full exposure of 1392 $\times$ 1040 12-bit pixels in 60\,ms.

To circumvent this long transfer time, the camera is operated in ``overlapped exposure mode." Here, the background exposure occurs while the signal exposure is uploaded. The background exposure cannot finish until the signal exposure has been completely transmitted. As a result, the background exposure must last at least 60\,ms. However, the background exposure can begin less than 0.5\,ms after the signal exposure ends.

The timing diagram of Fig.~\ref{fig:ImagingTimingDiagram} shows how overlapped exposure mode reduces the delay time. The signal image is acquired at the end of the first exposure while the background image is taken at the beginning of the second exposure. The images coincide with the opening of the second fiber shutter. To reduce stray light while the mechanical shutter is open, the first fiber shutter is opened 0.02\,ms before the second fiber shutter opens, and is closed 0.02\,ms after the second fiber shutter closes. The delay between images is so short that the atoms can appear in both images; to prevent this, the laser frequency is detuned by -15\,MHz in the interim.

Overlapped exposure mode only constrains the duration of the second exposure. To ensure that the camera dark current contributes equal background levels to the two exposures, both exposure durations should be set to 60\,ms. In this case, the first exposure begins (60\,ms - TOF) before the start of the imaging stage. When the first exposure begins during evaporation, Z-coil transport, or chip transfer, all of the lasers are shuttered, blocking stray laser photons from reaching the camera. However, when imaging atoms immediately after a stage when the lasers are not shuttered (e.g. compression or PGC), the camera will be exposed to scattered light. In this case, the first exposure must be shortened so that it begins immediately after the lasers are shuttered.

\section{Diagnostics and Calibration}

To evaluate vacuum performance, the 3D MOT lifetime and the magnetic trap lifetime were measured. In addition, to calibrate the atom number for absorption images, the image size per pixel was obtained by dropping clouds of atoms for various times-of-flights.

\begin{figure}[t]
\includegraphics[width=3.1in]{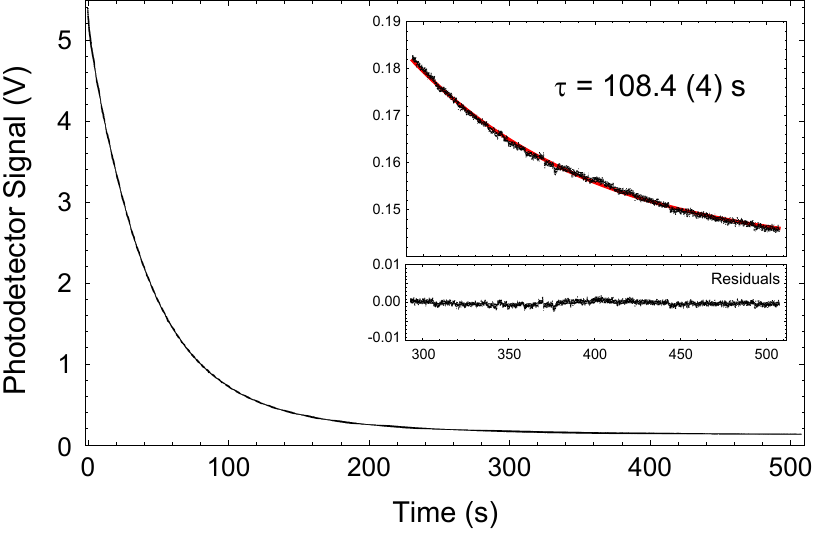}
\caption{The vacuum quality in the upper chamber of the vacuum system can be quantified by measuring the decay of a 3D MOT. For this data, the data points in the tail were fit to the sum of an exponential decay and constant offset using a Levenberg-Marquardt algorithm. Shown in red, the best-fit curve has a 1/$e$ time constant exceeding 100\,s (see inset).}
\label{fig:MOTDecay}
\end{figure}

\subsection{3D MOT Lifetime}

Decay of the 3D MOT is a useful diagnostic tool for verifying that the vacuum quality in the upper chamber is adequate for BEC production. Figure~\ref{fig:MOTDecay} shows a typical decay curve, as measured with the fluorescence photodetector in the physics package. At earlier times, decay is dominated by excited-state trap-loss collisions (i.e. fine-structure-changing collisions and radiative escape) that depend on the atom density, laser detuning, and laser intensity~\cite{Weiner1999}. In the tail of the decay, the atom density is low enough that loss is dominated instead by one-body collisions with the background gas. By fitting the tail to the sum of an exponential decay and constant offset, a time constant can be determined. For the data in Fig.~\ref{fig:MOTDecay}, this time constant exceeds 100\,s.

\subsection{Calibrating Atom Number}

The number of atoms $N$ obtained from an OD image is given by
\begin{eqnarray}
N = k^2\frac{1+S}{\sigma_0} \sum_i \text{OD}_i,
\label{eqn:AtomNumber}
\end{eqnarray}
where the sum is taken over pixels in the region of interest, $k^2$ is the area imaged by a single (square) camera pixel, $\sigma_0 = 0.1589$\,$\mu$m$^2$ is the resonant absorption cross-section for the $F=2 \rightarrow F'=3$ transition (assuming no magnetic field and a $\pi$-polarized laser beam), and $S = 0.15$ is the saturation parameter.

\begin{figure}[t]
\includegraphics[width=3.1in]{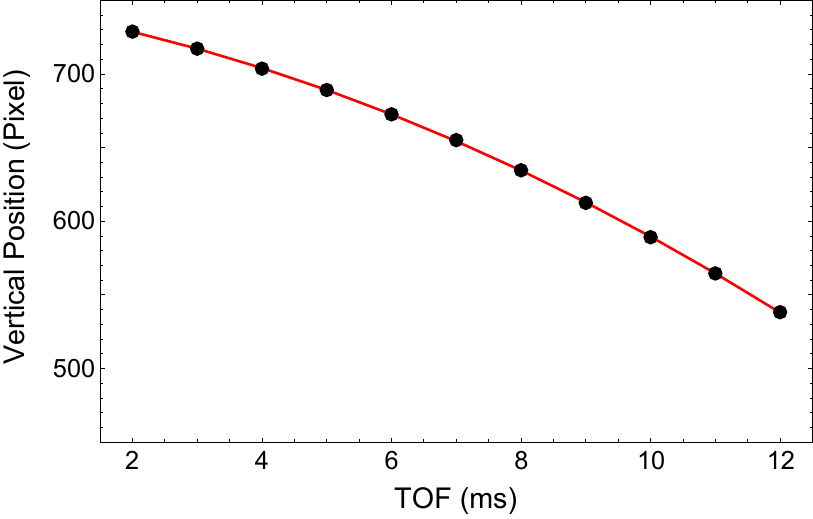}
\caption{After partial evaporation in the chip trap, thermal clouds were dropped for various times-of-flights (TOF). The vertical positions of these dropped clouds were fit to a parabola, from which the image size of a single CCD pixel was determined to be $k =$ 5.82(9)\,$\mu$m/pixel.}
\label{fig:PixelCalibration}
\end{figure}

To measure $k$, clouds were released from the same initial chip trap for various times-of-flights between 2 and 12\,ms. For each time-of-flight $t_i$, the OD image of the cloud was fit to a two-dimensional Gaussian, from which the center coordinates (in units of pixels) were extracted as best-fit parameters. When plotted versus time-of-flight, the vertical positions $y_i$ form a parabola due to the influence of gravity (see Fig.~\ref{fig:PixelCalibration}).

Quantitatively, the points $y_i(t_i)$ were fit to the function $y_i = 1/2 (g/k) t_i^2 + v_0 t_i + y_0$, where $g = 9.80$\,m/s$^2$, $v_0$ is a small initial downward velocity imparted to the cloud when it is released at $t=0$, and $y=0$ is the initial vertical position. To narrow the clouds' profiles and increase OD, the atoms were partially evaporated in the atom chip before dropping. The data fit returned the best-fit value $k = 5.82(9)\,\mu \text{m}/\text{pixel}$. Here, the choice of fitting function assumed that one of the camera's axes was aligned along the direction of gravity.

\subsection{Atom Chip Trap Lifetime}

\begin{figure}[t]
\includegraphics[width=3.1in]{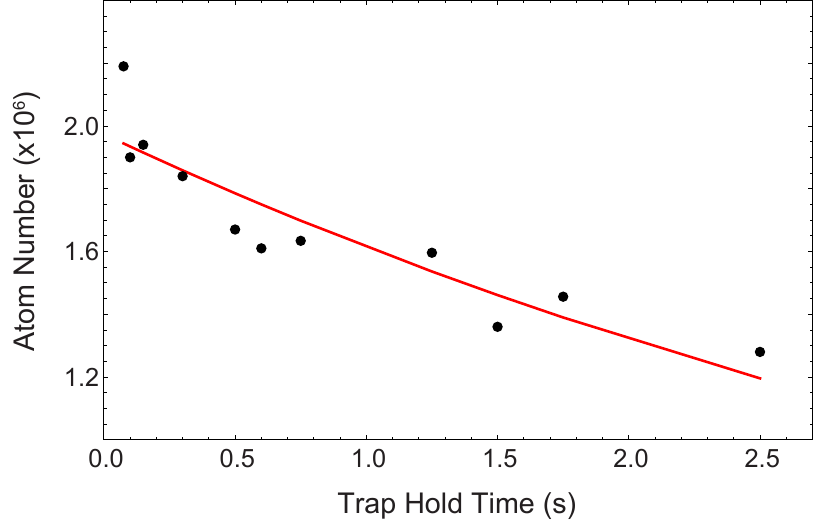}
\caption{Measured atom numbers as a function of hold time in a dimple trap. By fitting the atom numbers to an exponential function, a best-fit 1/$e$ trap lifetime of 5.0(8)\,s was obtained.}
\label{fig:ChipTrapLifetime}
\end{figure}

In addition to 3D MOT lifetime measurements, magnetic lifetime measurements are also useful for quantifying the residual background pressure in the upper chamber. As an example, Fig.~\ref{fig:ChipTrapLifetime} shows measured atom numbers as a function of hold time in a chip-based dimple trap. Here, atoms were first evaporatively cooled for 250\,ms, after which the trap was decompressed by reducing the bias fields. Each cloud was imaged after a 2\,ms time-of-flight. The atom numbers were determined from the OD images using Eqn.~\ref{eqn:AtomNumber}. The atom numbers were fit to an exponential function, from which a 1/$e$ lifetime of 5.0(8)\,s was obtained. Note that this lifetime is longer than the duration of the evaporative cooling sequence (see Table~\ref{tab:ExpParameters}), as required for runaway evaporation.

\section{Data Analysis}

To characterize the repeatability and stability of BEC production, principal component analysis (PCA) was applied to the images in Fig.~\ref{fig:BECmatrix}~\cite{Segal2010}. PCA can help identify the types and sources of noise and drift, and help distinguish fundamental noise processes (e.g. atom shot noise) from instrument-based systematics (e.g. temperature coefficients of electronic components). The main result of this analysis is an eigenvalue spectrum that quantifies the magnitude and distribution of variance in the data set. This spectrum can be used as a baseline to which other apparatus can be compared, or as an experimental ``signal" to be minimized as part of an optimization sequence.

\begin{figure}[t]
\includegraphics[width=3.25in]{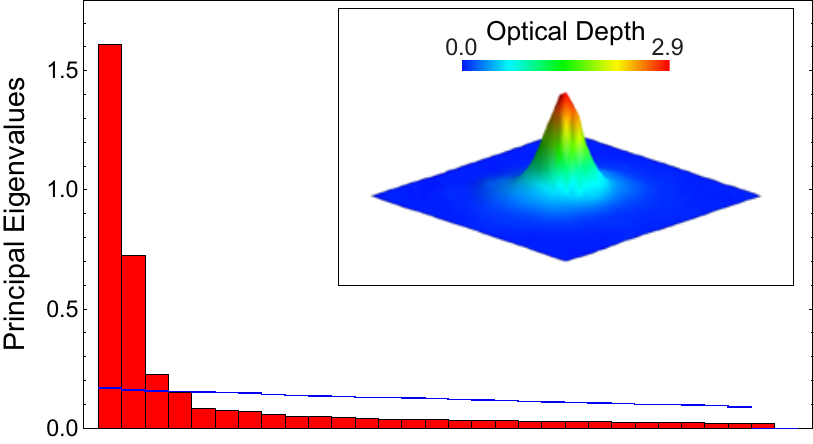}
\caption{Bar chart (red) of the ordered eigenvalues of the principal components of the data set in Fig.~\ref{fig:BECmatrix}. The first three principal components account for most of the variance of the data set. The blue lines show the eigenvalues of the principal components of white Gaussian noise whose total variance is equal to that of the data set. The image in the inset is comprised of the mean values of the pixels.}
\label{fig:PCABarChart}
\end{figure}

To perform PCA, the pixel values of each image were flattened to a one-dimensional array of length $m = 31 \times 31 = 961$. These arrays were used to construct a $m \times n$ matrix, where $n = 30$ was the number of samples for each pixel. For each of the $m$ pixels, the mean value of that pixel was subtracted from each of its samples. The image corresponding to the mean values of the pixels is shown in the inset of Fig.~\ref{fig:PCABarChart}.

The eigenvalues and eigenvectors of the covariance of this matrix were numerically calculated in Mathematica. The sorted eigenvalues are shown in red in Fig.~\ref{fig:PCABarChart}. Most of the variance of the data set can be attributed to the first three eigenvalues. For comparison, if all of the variance in the data set arose from independent, identically distributed (i.i.d.) noise taken from a probability distribution with a well-defined mean and variance (e.g. white Gaussian noise), then the eigenvalue spectrum would be represented by the blue lines in Fig.~\ref{fig:PCABarChart}. The distinct deviation from i.i.d. noise exhibited by the first few eigenvalues indicates the presence of systematic effects that may be non-uniform, either temporally or spatially.

\begin{figure*}[t]
\includegraphics[width=6.5in]{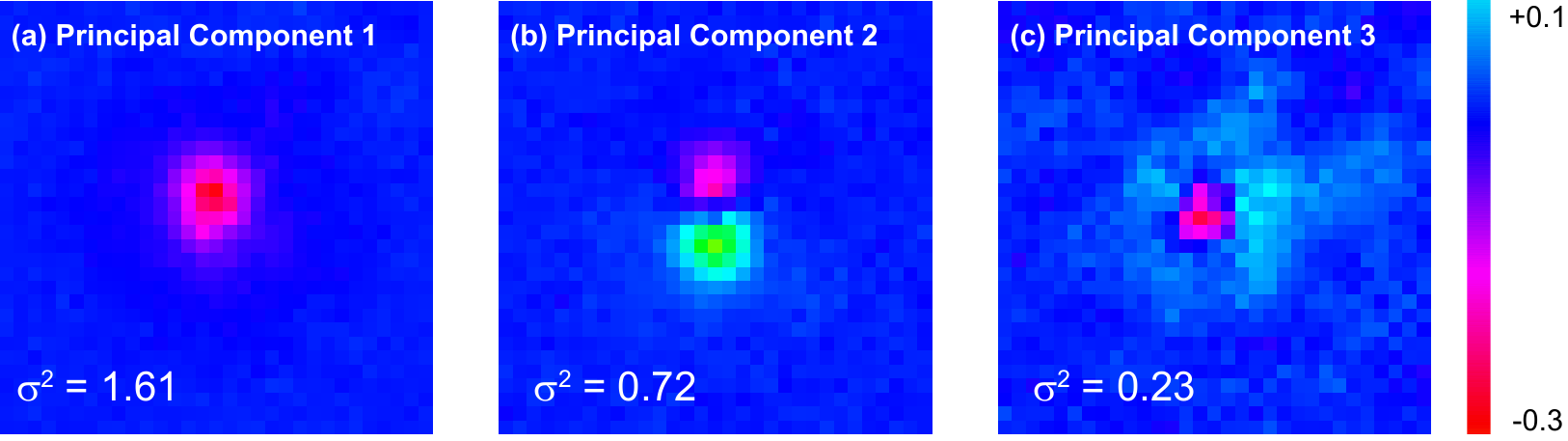}
\caption{The eigenvectors of the first three principal components show variances that are higher in the region of the cloud. In (a), the largest source of variance is most likely due to atom number fluctuations. In (b), the variance is attributed to fluctuations in the vertical position of the cloud. The principal component in (c) is interpreted as fluctuations in the condensate fraction.}
\label{fig:PCAimages}
\end{figure*}

To help identify processes that contribute to the eigenvalue spectrum, Fig.~\ref{fig:PCAimages} shows false-color plots of the first three eigenvectors. In Fig.~\ref{fig:PCAimages}a, the first eigenvector displays rotational symmetry about the location of the cloud's center. This eigenvector, which accounts for almost 40\% of the variance of the data set, is most likely due to atom number variation.

In Fig.~\ref{fig:PCAimages}b, the second eigenvector displays odd symmetry in the vertical direction about the position of the cloud's center, indicating fluctuations in the vertical position of the cloud. There is no principal component showing similar fluctuations in the horizontal direction. Possible explanations for this variation include timing jitter; changes in the chip trap's location due to noise and drift in the chip currents or magnetic bias fields; an initial downward velocity imparted to the cloud from an induced slosh mode; and mechanical vibrations of either the camera, imaging lens, or vacuum cell.

The third eigenvector, shown in Fig.~\ref{fig:PCAimages}c, contains a peak at the location of the cloud's center surrounded by a circular ring. This component is interpreted as fluctuations in the condensate fraction. Indeed, the BECs used in the data set were only partially condensed in the hopes of exaggerating such an effect.

To obtain values for atom number, condensate fraction, and cloud position, each cloud is modeled as a two-dimensional bimodal distribution that represents the condensed core as a parabaloid and the thermal component as a Bose-enhanced Gaussian~\cite{Castin1996,W.Ketterle1999}. Mathematically, this distribution $f(x,y)$ has the form
\begin{eqnarray}
f(x,y) = A^{(th)} \hspace{0.05in} g_2 \left\{ \exp \left[ - \bigg( \frac{x-x_0}{\sigma_x^{(th)}} \bigg)^2 -\bigg( \frac{y-y_0}{\sigma_y^{(th)} } \bigg)^2 \right] \right\}  \nonumber \\
+ \hspace{0.05in} A^{(c)} \hspace{0.05in}  \text{Max} \left[ 1-  \bigg( \frac{x-x_0}{\sigma_x^{(c)}} \bigg)^2 - \bigg( \frac{y-y_0}{\sigma_y^{(c)}} \bigg)^2 , \, 0   \right] + C \hspace{0.25in}
\label{eqn:fittingfunction}
\end{eqnarray}
where $g_j\{z\} = \sum_i z^i/i^j$,$\hspace{0.1in}x_0$ and $y_0$ are the center coordinates of the cloud, $\sigma_x^{(th)}$ and $\sigma_y^{(th)}$ are the widths of the thermal component along the $x$ and $y$ directions, $\sigma_x^{(c)}$ and $\sigma_y^{(c)}$ are the widths of the condensed component along the $x$ and $y$ directions, and $A^{(th)}$ and $A^{(c)}$ are the amplitudes of the thermal and condensed components, respectively. $C$ is a constant offset.

The condensate fraction $F = N_c / (N_c + N_{th})$ is obtained from the numbers $N_c$ and $N_{th}$ of condensed and thermal atoms, respectively. Integrating the first two terms of Eqn.~\ref{eqn:fittingfunction} over all space yields
\begin{eqnarray}
N_c = \frac{2\pi}{5} \frac{1+S}{\sigma_0} A^{(c)} \sigma_x^{(c)} \sigma_y^{(c)}
\label{eqn:Nc}
\end{eqnarray}
where $\sigma_0 = 0.1589\,\mu$m$^2$ is the resonant absorption cross section for the $F=2 \rightarrow F'=3$ transition (for no magnetic field and $\pi$-polarized laser light) and $S=0.15$ is the saturation parameter. Similarly,
\begin{eqnarray}
N_{th} = g_3\{1\}\,\pi \frac{1+S}{\sigma_0} A^{(th)} \sigma_x^{(th)} \sigma_y^{(th)}
\label{eqn:Nth}
\end{eqnarray}
where $g_3\{1\} = \zeta(3) \approx 1.202$. Here, the widths $\sigma_x$ and $\sigma_y$ are expressed in units of meters by multiplying their values in pixels by the calibration factor $k = 5.82\,\mu \text{m}/\text{pixel}$.

\begin{figure}[b]
\includegraphics[width=3.25in]{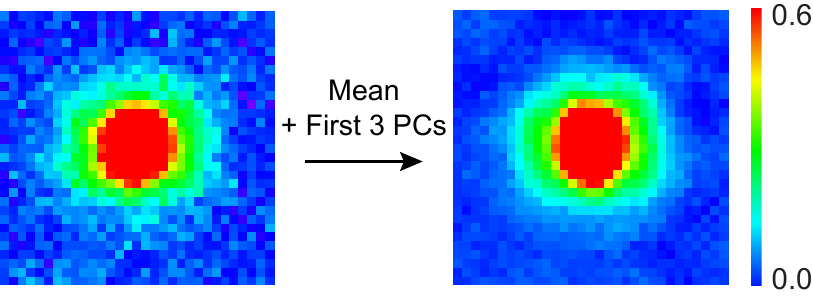}
\caption{Example of image filtering based on PCA. Here, the first BEC in Fig.~\ref{fig:BECmatrix} was reconstructed using the mean image (the inset of Fig.~\ref{fig:PCABarChart}) and the first three principal components.}
\label{fig:FilteredImage}
\end{figure}

\begin{figure}[b!]
\includegraphics[width=2.8in]{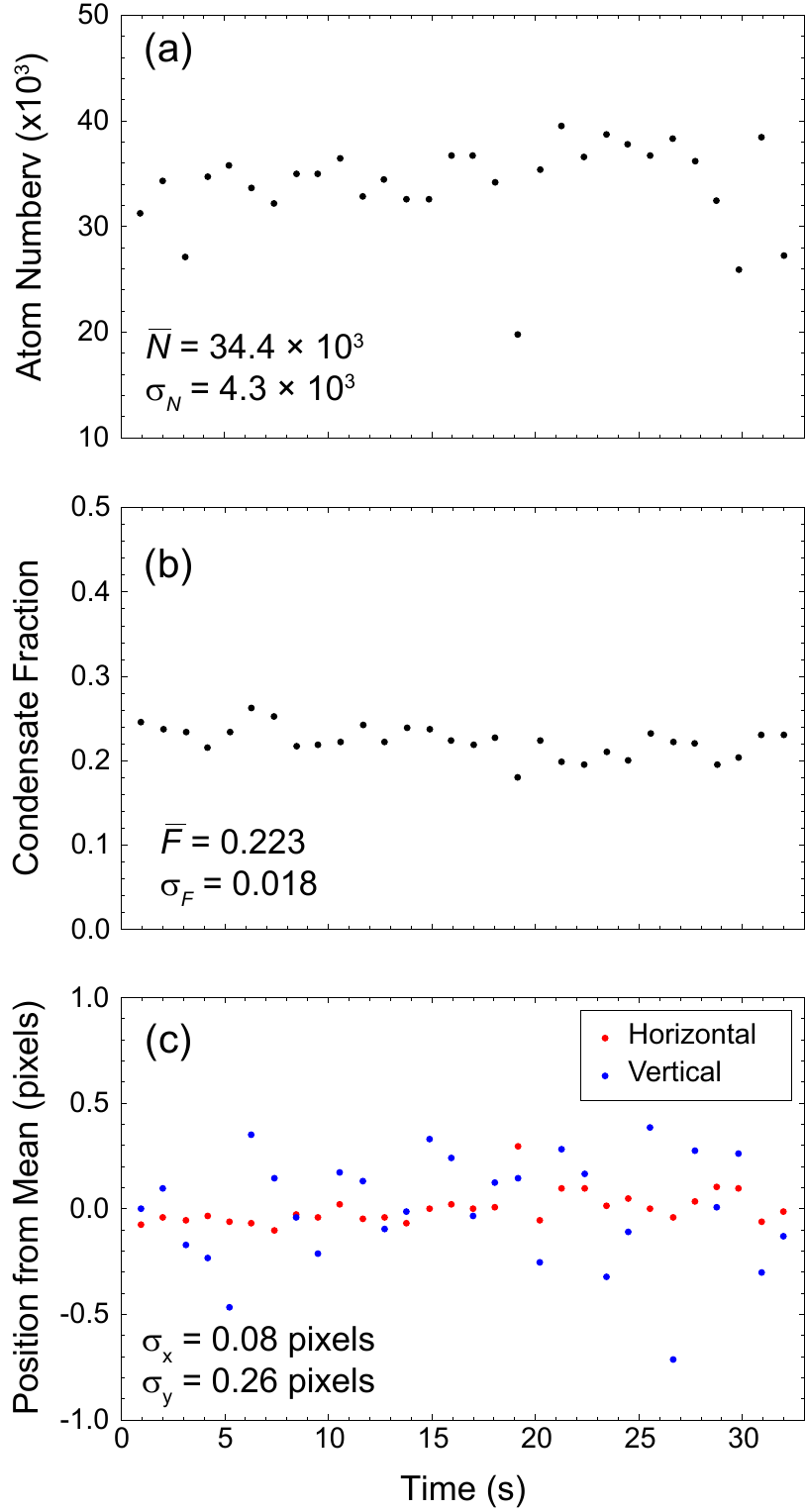}
\caption{Total atom numbers, condensate fractions, and cloud positions were obtained by fitting each filtered BEC image to a bimodal distribution (see Eqn.\,\ref{eqn:fittingfunction}). The vertical positions have a standard deviation $\sigma_y$ three times larger than that for the horizontal positions ($\sigma_x$).}
\label{fig:BestFitPlots}
\end{figure}

To filter the images in Fig.~\ref{fig:BECmatrix}, each of the 30 images was reconstructed using the mean image and the first three PCs. Keeping the PCs with the largest eigenvalues ensures that the filtered images retain the key sources of variance. The other 27 PCs predominantly reflect i.i.d. noise that is more likely due to imaging noise than any true variation of the atom clouds. Therefore, discarding these less significant PCs yields cleaner images while having negligible impact on the ability to identify and characterize the dominant systematic effects. As an example of filtering, Fig.~\ref{fig:FilteredImage} shows the first BEC image both before and after filtering.

Each of the filtered BEC images was fit to Eqn.~\ref{eqn:fittingfunction} using a Levenberg-Marquardt algorithm. Using the best-fit parameters, the cloud position $(x_0,y_0)$, total atom number $N=N_c + N_{th}$, and condensate fraction $F$ were calculated for each image. Figures~\ref{fig:BestFitPlots}a and~\ref{fig:BestFitPlots}b show the total atom numbers and condensate fractions, respectively. For atom number, the mean is $\bar{N}=34.4~\times~10^3$ atoms with a standard deviation of 4.3$~\times~10^3$ atoms. For condensate fraction, the mean is $\bar{F}=0.223$ with a standard deviation of 0.018.

For the total atom numbers and condensate fractions shown in Fig.~\ref{fig:BestFitPlots}, the mean number of condensed atoms is $\bar{N}_c=7.7~\times~10^3$. However, when producing the BECs in Fig.~\ref{fig:BECmatrix}, the lower frequency of evaporation stage E was slightly raised to inhibit complete condensation. By using partially condensed clouds, it was easier to obtain condensate fractions from the data analysis, and therefore to use this parameter to help characterize the stability of BEC production. When the final rf frequency in this last evaporation stage was lowered by a few kilohertz, nearly pure BECs with 1.5$~\times~10^4$ atoms were produced. This latter value is shown in Fig.~\ref{fig:BECmachines}.

Fig.~\ref{fig:BestFitPlots}c shows the best-fit positions for the cloud centers. As expected, the standard deviation of 0.25 pixels in the vertical direction (blue points) is three times larger than the standard deviation of 0.07 pixels in the horizontal direction. Note that the variance in the vertical position appears to arise from an increase in shot-to-shot scatter, as opposed to a drift on longer timescales. This feature indicates that slow processes, such as temperature drift, are unlikely to explain the variance.

\section{Conclusions}

We presented an ultracold-atom system that is capable of producing $^{87}$Rb BECs at a rate of 1\,Hz. By confining the atoms in a tight dimple trap created with an atom chip, the duration of evaporative cooling was reduced to less than 0.5\,s. The system can operate continuously, as demonstrated by the sequential creation and imaging of 30 BECs (see Fig.~\ref{fig:BECmatrix}). Principal component analysis was used to study the variance of the BEC images and to provide insight into some of the dominant sources of noise and drift. The system's speed, repeatability, and size make it a useful starting point for a portable ultracold-atom system designed for inertial sensing, gravimetry, and other applications outside of a laboratory environment.

\pagebreak

\begin{acknowledgments}
This work was funded by the Office of Naval Research through their Small Business Innovation Research (SBIR) program (Contract Number N00014-10-C-0282).
\end{acknowledgments}

\bibliography{FastBECrefs}

\end{document}